\documentclass{aa}  

\usepackage{graphicx}
\usepackage{txfonts}
\usepackage{color}

\begin{document}

   \title{Asymmetries on red giant branch surfaces from CHARA/MIRC optical interferometry}
\titlerunning{Asymmetries on red giant branch surfaces from CHARA/MIRC optical interferometry}
%     \subtitle{using MIRC data}

  \author{A. Chiavassa \inst{1}, R. Norris \inst{2}, M. Montarg\`es \inst{3}, R. Ligi \inst{4}, L. Fossati \inst{5}, L. Bigot \inst{1}, F. Baron \inst{2}, P. Kervella \inst{6,7}, J. D. Monnier \inst{8}, D. Mourard \inst{1}, N. Nardetto \inst{1}, G. Perrin \inst{8}, G.~H. Schaefer\inst{9}, T.~A. ten Brummelaar\inst{9}, Z. Magic\inst{10,11}, R. Collet\inst{12}, M. Asplund\inst{13}}
  %\authorrunning{A. Chiavassa et al.}
\authorrunning{A. Chiavassa et al.}
\institute{Universit\'e C\^ote d'Azur, Observatoire de la C\^ote d'Azur, CNRS, Lagrange, CS 34229, Nice,  France \\
\email{andrea.chiavassa@oca.eu}
\and
CHARA and Department of Physics $\&$ Astronomy, Georgia State University, P. O. Box 4106, Atlanta, GA 30302-4106, USA
\and 
Institut de Radioastronomie Millim\'etrique, 300 rue de la Piscine, 38406, Saint Martin d'H\`eres, France
\and
Aix Marseille Universit\'e, CNRS, LAM (Laboratoire d'Astrophysique de Marseille) UMR 7326, 13388, Marseille, France
\and
Space Research Institute, Austrian Academy of Sciences, Schmiedlstrasse 6, A-8042 Graz, Austria
\and
Unidad Mixta Internacional Franco-Chilena de Astronom\'ia (UMI 3386), CNRS/INSU, France $\&$ Departamento de
Astronom\'ia, Universidad de Chile, Camino El Observatorio 1515, Las Condes, Santiago, Chile
\and
LESIA, Observatoire de Paris, PSL Research University, CNRS UMR 8109, Sorbonne Universit\'es, UPMC, Universit\'e
Paris Diderot, Sorbonne Paris Cit\'e, 5 place Jules Janssen, F-92195 Meudon, France
\and
Department of Astronomy, University of Michigan, 918 Dennison Building, Ann Arbor, MI48109-1090, USA
\and
The CHARA Array of Georgia State University, Mount Wilson, CA 91023, USA
\and
Niels Bohr Institute, University of Copenhagen, Juliane Maries Vej 30, DK--2100 Copenhagen, Denmark  
\and 
Centre for Star and Planet Formation, Natural History Museum of Denmark, University of Copenhagen, {\O}ster Voldgade 5-7, DK--1350 Copenhagen, Denmark
\and
Stellar Astrophysics Centre, Department of Physics and Astronomy, Ny Munkegade 120,  Aarhus University, DK-8000 Aarhus C, Denmark
\and
Research School of Astronomy $\&$ Astrophysics, Australian National University, Cotter Road, Weston ACT 2611, Australia}
\date{...; ...}

% \abstract{}{}{}{}{} 
% 5 {} token are mandatory
 
  \abstract
    {Red giant branch (RGB) stars are very bright objects in galaxies and are often used as standard candles. Interferometry is the ideal tool to characterize the dynamics and morphology of their atmospheres.}
   {We aim at precisely characterising the surface dynamics of a sample of RGB stars.}
   {We obtained interferometric observations for three RGB stars with the MIRC instrument mounted at the CHARA interferometer. We looked for asymmetries on the stellar surfaces using limb-darkening models.}
   {We measured the apparent diameters of HD197989 ($\epsilon$ Cyg) = 4.61$\pm$0.02 mas, HD189276 (HR7633) = 2.95$\pm$0.01 mas, and HD161096 ($\beta$ Oph) = 4.43$\pm$0.01 mas. We detected departures from the centrosymmetric case for all three stars with the tendency of a greater effect for lower log$g$ of the sample. We explored the causes of this signal and conclude that a possible explanation to the interferometric signal is the convection-related and/or the magnetic-related surface activity. However, it is necessary to monitor these stars with new observations, possibly coupled with spectroscopy, in order to firmly establish the cause.}
   {}


   \maketitle
%
%-------------------------------------------------------------------

\section{Introduction}

Red giant branch (RGB) stars have evolved from the main sequence and are powered by hydrogen burning in a thin shell surrounding their helium core. This evolutionary phase precedes the asymptotic giant branch (AGB) and is characterised during the evolved states by more expanded and deformed outer layers. RGB stars are bright candles in galaxies, and the accurate determination of their fundamental parameters and chemical composition is essential for tracing the morphology and the evolution of the Galaxy, for probing distant stellar populations, and for characterising globular clusters. Their masses are typically lower than $\sim$2.0
$M_\odot$ \citep{2002PASP..114..375S} with $4\,000\lesssim T_{\rm{eff}}\lesssim5\,100$ K depending on metallicity, $1.5\lesssim\log{g}\lesssim3.5$, and $3\lesssim R_\star\lesssim70$ $R_\odot$  \citep{1999AJ....117..521V,2010ApJ...710.1365B}.

Like all late-type stars, red giant atmospheres are made complex by convective motions and turbulent flows. Convection contributes significantly to the transportation of energy from the stellar interior to the outer layer, and in the photosphere, it manifests itself as a granulation pattern characterised by dark intergranular lanes of downflowing cooler plasma and bright areas (the granules) where hot plasma rises \citep[see the review of][]{2009LRSP....6....2N}. The granules cause an inhomogeneous stellar surface that changes with time. The granulation potentially acts as an intrinsic noise in stellar parameters, radial velocity, and chemical abundance determinations. In addition to this, a magnetic field may be present, as detected in several RGB stars by \cite{2015A&A...574A..90A}, and this may contribute to a bias in the stellar parameter determination. 
The characterisation of the dynamics and morphology of RGB stars is important to quantify the  effect of the granulation and magnetic fields, and thanks to its high angular resolution, interferometry is the ideal tool for this purpose.

In this Letter, we present the detection of an interferometric signal at high spatial frequencies for three RGB stars using the MIRC instrument mounted at the CHARA interferometer. We analyse the possible causes of this signal.

\begin{table}
\scriptsize
        \caption{Log of observations}              % title of Table
        \label{log}      % is used to refer this table in the text
        \centering                                      % used for centering table
        \begin{tabular}{c c c c c}          % centered columns (4 columns)
                \hline\hline                        % inserts double horizontal lines
                Date (UT) & Target & $N_{block}$ & TTO   & Calibrators \\    % table heading
                                  &        &             &  (min) &                \\
                \hline                                    % inserts single horizontal line
                2016 Jul 6 & HD161096 ($\beta$ Oph) & 2 & 20 & $\epsilon$ Ser, 72 Oph,$\lambda$ Aql  \\
                2016 Jul 6 & HD189276 (HR 7633) & 1 & 10 & $\iota$ Cyg, HR 8185 \\
                2016 Jul 6 & HD197989 ($\epsilon$ Cyg) & 1 & 15 & $\gamma$ Lyr\\
                2016 Jul 7 & $\beta$ Oph & 1 & 25 & $\epsilon$ Ser, 72 Oph\\
                2016 Jul 7 & HR 7633 & 1 & 20 & $\iota$ Cyg, \\
                2016 Jul 7 & $\epsilon$ Cyg & 1 & 30 & 17 Cyg, $\sigma$ Cyg\\
                \hline
        \end{tabular}
        \tablefoot{TTO is the total time observed. \\
                Calibrator diameters (mas): $\epsilon$ Ser = 0.689 $\pm$ 0.048 (1)\\ 72 Oph = 0.684 $\pm$ 0.048 (2); $\lambda$ Aql = 0.518 $\pm$ 0.036 (2) \\ $\iota$ Cyg = 0.586 $\pm$ 0.041 (2); HR 8185 = 1.067 $\pm$ 0.076 (2) \\ $\gamma$ Lyr = 0.737 $\pm$ 0.015 (3); 17 Cyg = 0.721 $\pm$ 0.051 (2) \\ $\sigma$ Cyg = 0.542 $\pm$ 0.021 (4). All the diameters, except for $\epsilon$ Ser, are in the H band. $\epsilon$ Ser is given in K band, variation across wavelength for this diameter is extimated to be $\sim$0.01 mas with negligible effect on the data reduction.}\\
        \tablebib{
                (1)~\citet{Boyajian2012}; (2) \citet{searchcal}; (3) \citet{monnier_resolving_2012}; (4) \citet{2008ApJ...684L..95Z}}
\end{table}

%\begin{table}
%       \centering
%       \caption{Adopted calibrators and their stellar diameters.}
%       \label{log}
%       \begin{tabular}{l l l l}
%               \hline\hline\noalign{\smallskip}
%               Star & Calibrators & Diameter (mas) & Ref. \\
%               \hline\noalign{\smallskip}
%               $\epsilon$ Cyg & $\gamma$ Lyr & $0.737 \pm 0.015$ & 1\\
%               & $\sigma$ Cyg & $0.542 \pm 0.021$ & 2\\
%               & 17 Cyg & $0.721 \pm 0.051$ & 3\\
%               HR7633 & $\iota$ Cyg & $0.586 \pm 0.041$ & 3\\
%               & HR8185 & $1.067 \pm 0.076$ & 3\\
%               $\beta$ Oph & $\epsilon$ Ser & 0.689 $\pm$ 0.048 & 4 \\
%               & 72 Oph & 0.684 $\pm$ 0.048 & 3 \\
%               & $\lambda$ Aql &  0.518 $\pm$ 0.036 & 3\\
%               \hline
%       \end{tabular}
%\tablebib{(1) \cite{monnier_resolving_2012}; (2) %\cite{2008ApJ...684L..95Z}; (3) \citet{searchcal}; (4) \citet{Boyajian2012}}
%\end{table}

\begin{table*}
        %\begin{minipage}[t]{\columnwidth}
        \caption{Parameters of the RGB stars.}             % title of Table
        \label{startable}      % is used to refer this table in the text
        \centering                          % used for centering table
        \renewcommand{\footnoterule}{} 
        \begin{tabular}{c  c c  c c c c c }        % centered columns (4 columns)
                \hline\hline                 % inserts double horizontal lines
                Stars & Spectral  &  H mag  &  $M$         &  [Fe/H]  & $T_{\rm{eff}}$ &   log$g$   & R      \\
                & type\tablefootmark{a}&    & $[M_\odot]$ &          &  [K]                      &   [cgs]         &  $[R_\odot]$            \\
                \hline
                 $\epsilon$ Cyg & K0III-IV & 0.200\tablefootmark{b} &  1.84$\pm$0.31\tablefootmark{d}  & -0.11$\pm$0.10\tablefootmark{d}& 4778$\pm$49\tablefootmark{d} & 2.62$\pm$0.10\tablefootmark{d} & 11.08$\pm$0.25\tablefootmark{d}       \\
                 HR7633 & K4.5IIIa & 0.919\tablefootmark{b}  & -- & -- & 4050\tablefootmark{e} & 1.70\tablefootmark{e} & -- \\
                 $\beta$ Oph  & K2III & 0.354\tablefootmark{c} & 1.63$\pm$0.18\tablefootmark{d}  & 0.13$\pm$0.10\tablefootmark{d} & 4520$\pm$44\tablefootmark{d} & 2.42$\pm$0.07\tablefootmark{d} & 13.13$\pm$0.32\tablefootmark{d}       \\
                \hline                        % inserts single horizontal line
                
                \hline                                   %inserts single line
        \end{tabular}
        %\end{minipage}
        \tablefoot{
                \tablefoottext{a}{\cite{2003AJ....126.2048G}}
                \tablefoottext{b}{\cite{2002yCat.2237....0D}}
                \tablefoottext{c}{\cite{2012MNRAS.419.1637L}}
                \tablefoottext{d}{\cite{2015A&A...574A.116R}}
                \tablefoottext{e}{\cite{2010SPIE.7734E..4EL}}
        }
\end{table*}

\begin{figure}
        \centering
        \begin{tabular}{c}                         
                \includegraphics[width=0.95\hsize]{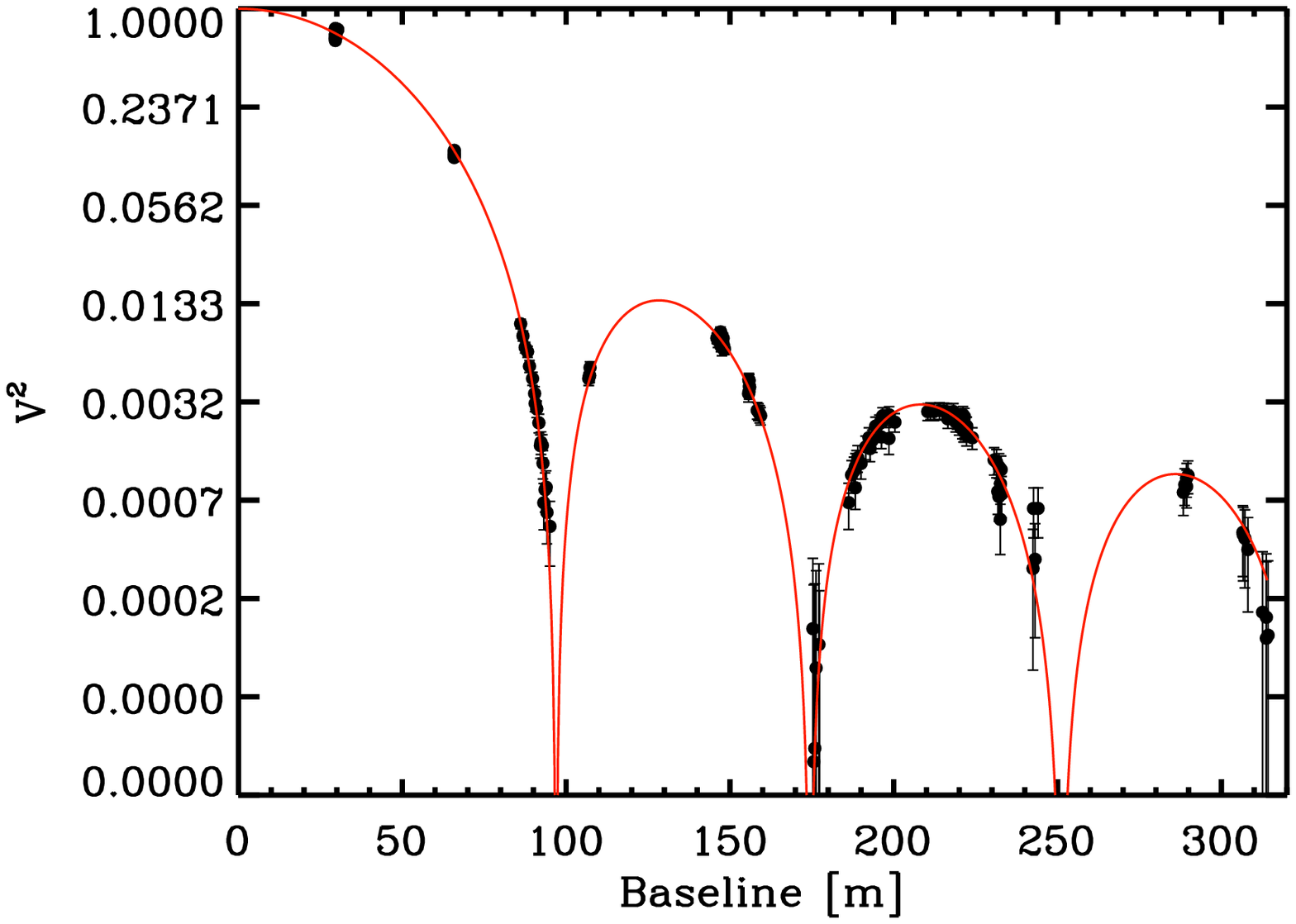}    \\
                \includegraphics[width=0.935\hsize]{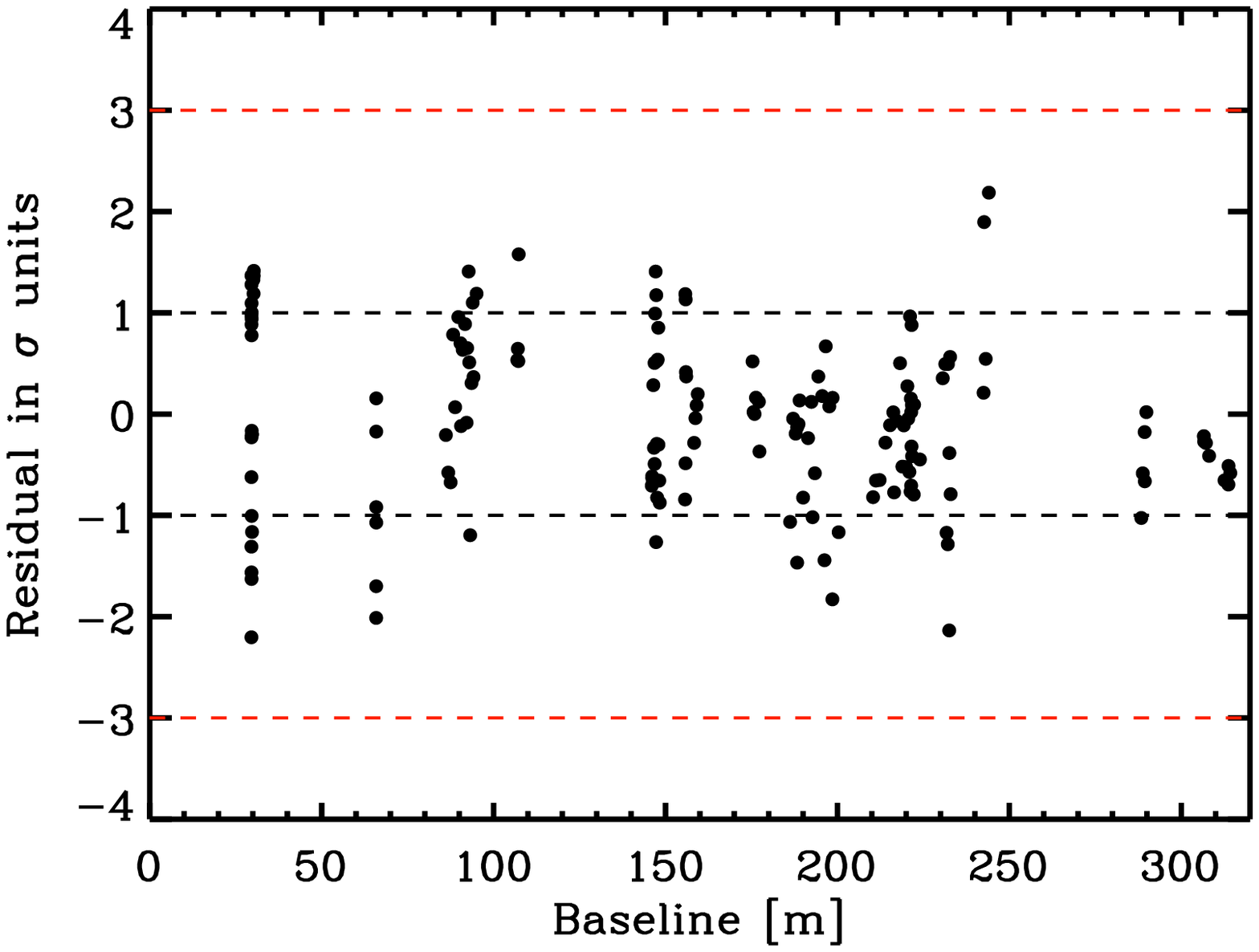}    \\
                \includegraphics[width=0.95\hsize]{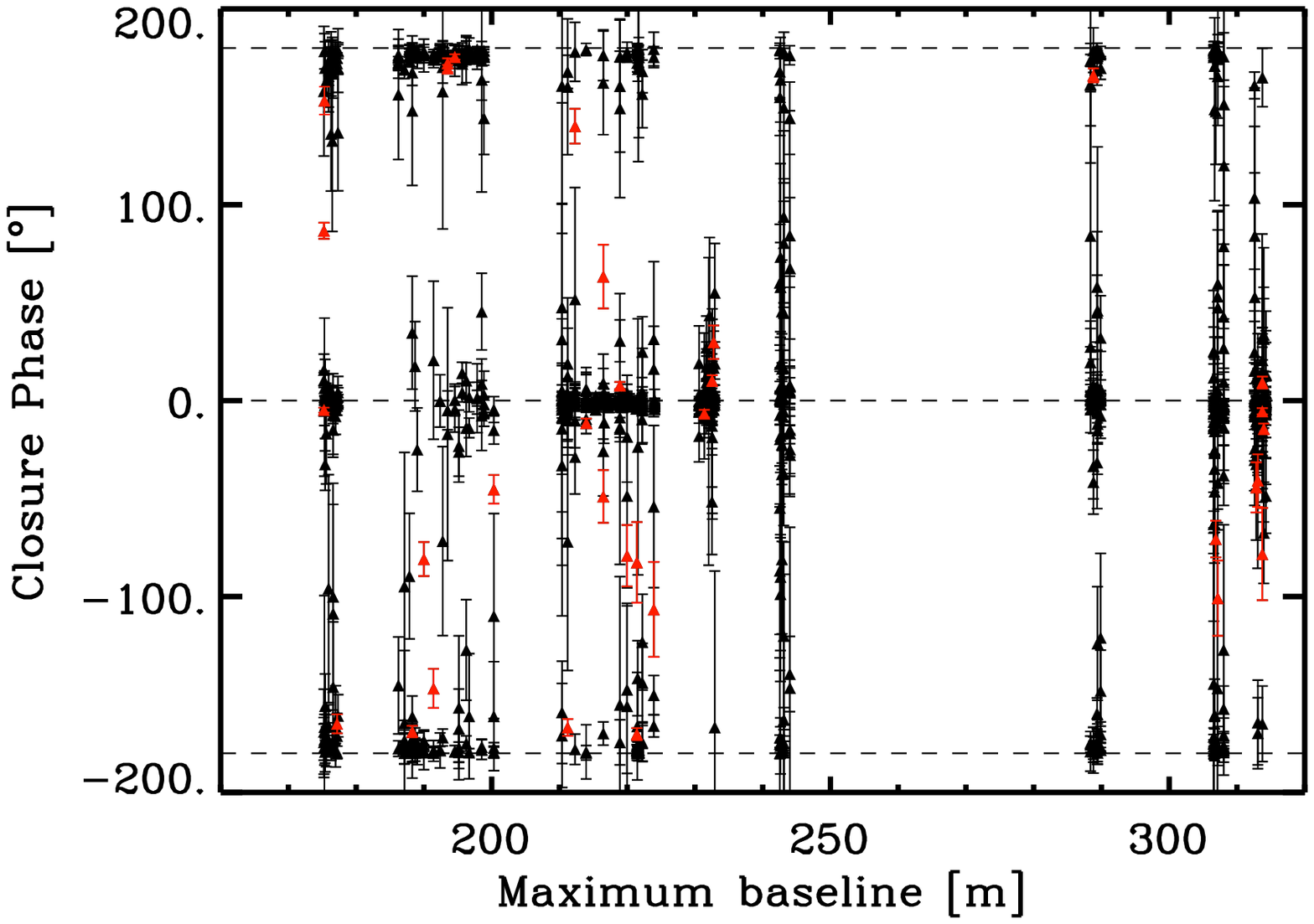}       
        \end{tabular}
        \caption{\emph{Top and central panels:} squared visibilities with the limb-darkening fit residuals (top and central panels) at the wavelength band 1.6004$\pm$0.0036 $\mu$m. The continuous line in the top panel corresponds to the limb-darkening fit whose parameters are reported in Table~\ref{phase}.   The black horizontal dashed lines in the central panel correspond to the value of 1$\sigma,$ and the red line shows the value of 3$\sigma$. \emph{Bottom panel:} closure phase data points of $\beta$ Oph (Table~\ref{startable}) for all the wavelengths. The red data correspond to closure phase departures larger than 3$\sigma$ (see Fig.~\ref{departures}). The horizontal dashed lines in the bottom panel display the zero or $\pm180^{\circ}$ values.} 
        \label{betoph}
\end{figure}

%--------------------------------------------------------------------
\section{Interferometric observations with MIRC at CHARA}

We collected observations of three stars (Table~\ref{log}) using the Michigan Infrared Combiner (MIRC) on the Georgia State University Center for High Angular Resolution Astronomy (CHARA). The CHARA array is located on Mount Wilson, CA, and consists of six 1 m
telescopes for a total of 15 baselines ranging in length from 34 m to
331 m, resulting in an angular resolution of $\sim$0.5 mas in the \emph
{H} band \citep{ten_brummelaar_first_2005}. The detailed parameters of the RGB stars are reported in Table~\ref{startable}.

The MIRC \citep{monnier_michigan_2004,monnier_resolving_2012} is a six-beam combiner operating in the \emph {H} band (1.5-1.8 $\mu$m) at low spectral resolution (R=30). Each observing block consists of observations of a calibrator, a target, and when possible, a second calibrator. Time spent collecting data on the target ranged within 10 to 30 minutes, excluding background and other calibration frames, depending on observing conditions. We used the latest version of the MIRC reduction pipeline (as of October 2016) and as previously described in \citet{2007Sci...317..342M} and \citet{monnier_resolving_2012}. The pipeline uses Fourier transforms to compute squared visibilities, which are then averaged and corrected for biases. We determined the bispectrum using the phases and amplitudes of three baselines in a closed triangle. We calibrated photometric amplitudes using a beam splitter following spatial filtering \citep{che_high_2010}\footnote{This research has made use of the Jean-Marie Mariotti Center \texttt{SearchCal} service  \cite{searchcal}, available at http://www.jmmc.fr/searchcal,  co-developped by FIZEAU and LAOG/IPAG, and of CDS Astronomical Databases SIMBAD and VIZIER, available at http://cdsweb.u-strasbg.fr/}. Because we do not expect significant brightness variation over short time periods for these targets, we combined the two nights of observations for each star into single files, accounting for systematic error by applying multiplicative and additive errors as described in \citet{monnier_resolving_2012}. At low visibilities  ($\lesssim 10^{-3}$), the signal-to-noise ratio of the data decreases because of cross-talk. We therefore remain cautious when interpreting data at such low visibilities.\\
The observations were collected in eight different wavelength bands: 1.7379$\pm$0.0031, 1.7055$\pm$0.0033, 1.6711$\pm$0.0035, 1.6361$\pm$0.0035, 1.6004$\pm$0.0036, 1.5642$\pm$0.0037, 1.5273$\pm$0.0035,
and 1.4833$\pm$0.0033 $\mu$m. In the following, we use the wavelength band centred at 1.6004$\pm$0.0036 $\mu$m for the apparent diameter determination with visibility curves because it corresponds to the H$^-$ continuous opacity minimum (consequently closest to the continuum forming region). For all the closure phase plots, we used the full set of wavelength bands.

\begin{table}
        \small
        \begin{minipage}[t]{\columnwidth}
                \caption{Apparent diameters of the observed stars at 1.6004$\pm$0.0036 $\mu$m}            % title of Table
                \label{phase}      % is used to refer this table in the text
                \centering                          % used for centreing table
                \renewcommand{\footnoterule}{} 
                \begin{tabular}{c c c c c }        % centreed columns (4 columns)
                        \hline\hline                 % inserts double horizontal lines
                        Star    & LD power-law&  $\theta_{LD}$ &  $\theta_{LD}$ & $\overline{\chi^2}$  \\
                        &   exponent  & [mas]    &  $[R_\odot]$  &  \\
                        \hline
                        $\epsilon$ Cyg &  0.25$\pm$0.01   & 4.61$\pm$0.02 &  22.09$\pm$0.15 & 2.45 \\
                        HR7633 & 0.14$\pm$0.01   & 2.95$\pm$0.01 &  178.11$\pm$11.11 & 1.32 \\
                        $\beta$ Oph &  0.25$\pm$0.01  & 4.43$\pm$0.01 &  23.89$\pm$0.16  & 1.01 \\
                        \hline\hline                          % inserts single horizontal line
                \end{tabular}
        \end{minipage}
\end{table}

\begin{figure*}
        \centering
        \begin{tabular}{ccc}                         
                \includegraphics[width=0.328\hsize]{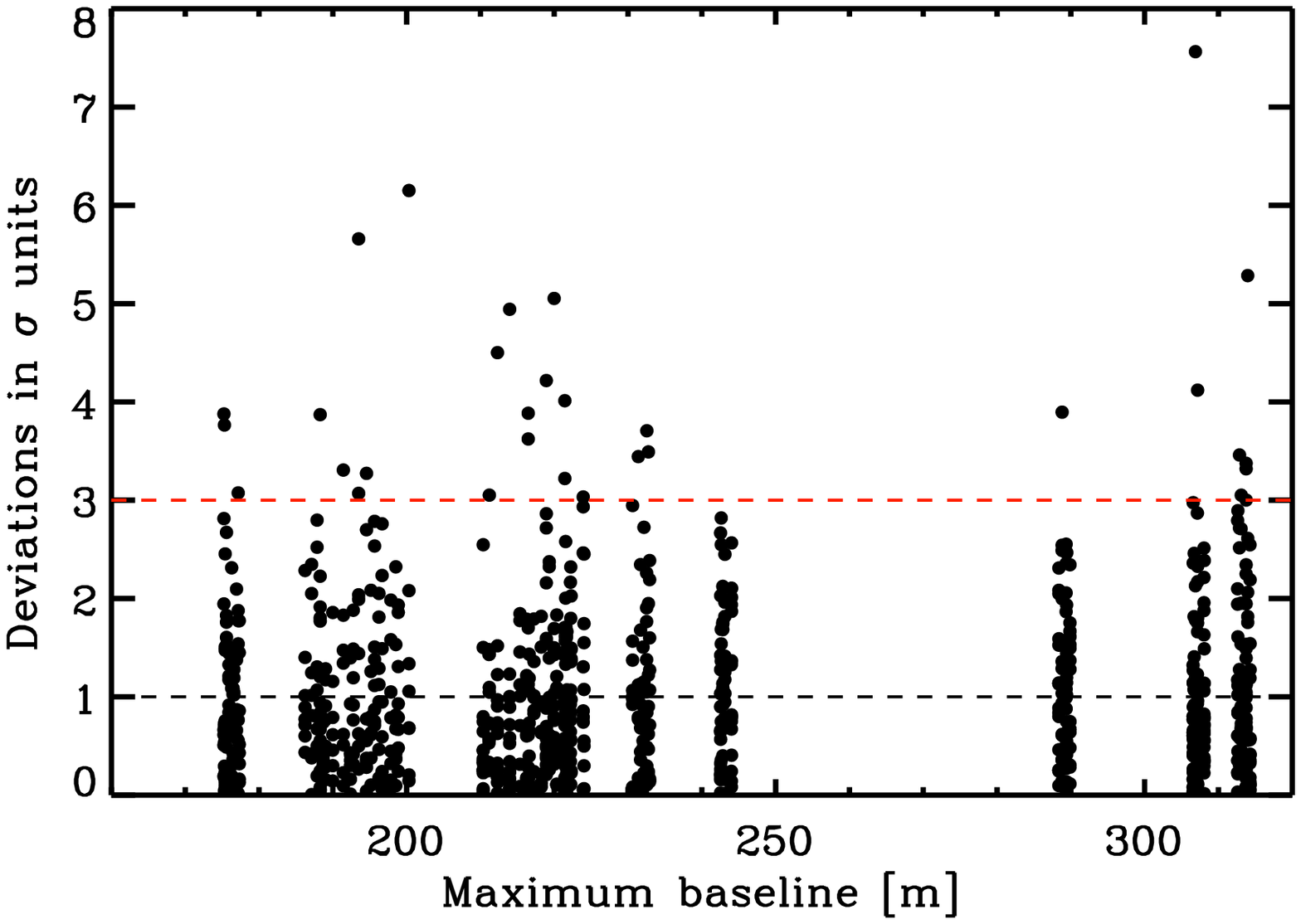}    
                \includegraphics[width=0.328\hsize]{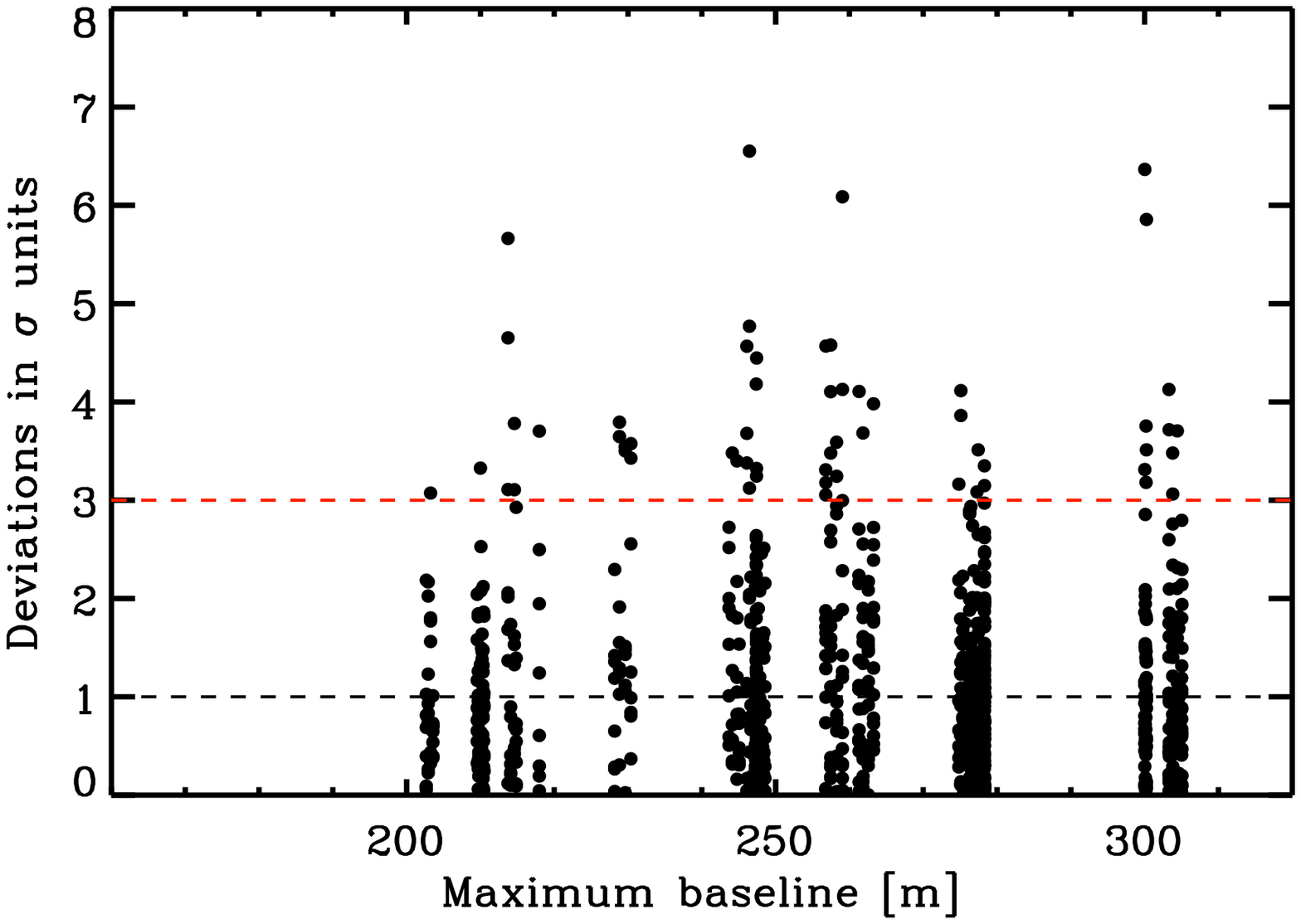}    
                \includegraphics[width=0.328\hsize]{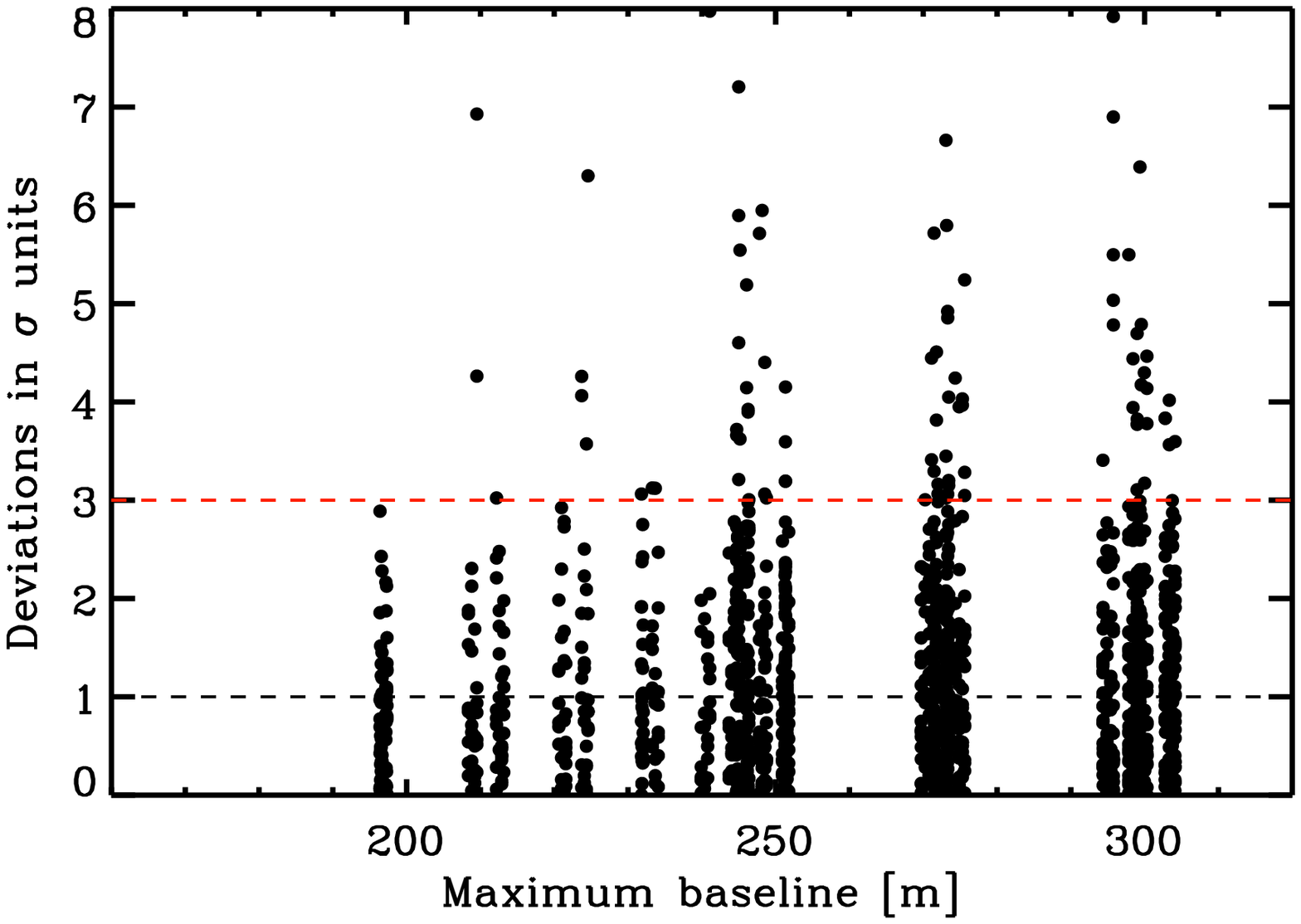}       
        \end{tabular}
        \caption{Closure phase departures from zero or $\pm180^{\circ}$ for all the observed stars and for all the wavelengths. For every data point, the lowest value between $\mid data-0^{\circ}\mid, \mid data-180^{\circ}\mid,$ and $\mid data+180^{\circ}\mid$ is computed and normalised by the corresponding observed error, $\sigma$. $\beta$ Oph is shown in the left panel, $\epsilon$ Cyg in the central, and HR7633 in the right panel. The horizontal black line corresponds to the value of 1$\sigma$ and the red line to 3$\sigma$.} 
        \label{departures}
\end{figure*} 

\section{Discussion}

%\subsection{Squared visibility and closure phase and departures from centrosymmetric case}

We fitted the data, based only on the squared visibilities, with an power-law limb-darkened disk model whose parametric values are reported in Table~\ref{phase}.  Figure~\ref{betoph} displays the example of $\beta$ Oph, while the other stars are reported in Figs.~\ref{epscyg} and \ref{HR7633}. The limb-darkening fit is very good (Table~\ref{phase}) with larger residuals for HR7633.
We report the first measure of the radius for HR7633, and while the radius of $\epsilon$ Cyg is in good agreement with \cite{2015A&A...574A.116R}, the radius of $\beta$ Oph is slightly smaller.\\
The observed closure phases display values different from zero or $\pm180^{\circ}$ for all the observed stars. To determine the amplitude of these deviations, we selected the lowest value between $\mid data-0^{\circ}\mid, \mid data-180^{\circ}\mid,$ and $\mid data+180^{\circ}\mid$ for each data point and then normalised it by the corresponding observed error, $\sigma$. We plot the data departures in Fig.~\ref{departures},  which shows several points diverging from the centrosymmetric case for values higher than 3 $\sigma$: the closure phase departures are smaller for $\beta$ Oph (33 points over 920, 3.6$\%$, higher than 3$\sigma$), intermediate for $\epsilon$ Cyg (70 points over 1056, 6.6$\%$, higher than 3$\sigma$), and larger for HR7633 (88 points over 1384, 6.3$\%$, higher than 3$\sigma$). For $\beta$ Oph and $\epsilon$ Cyg, the spatial frequencies spanned extend to the fourth lobe, while for HR7633, they only reach (partially) the third lobe. The contribution of the small-scale structures increases with the frequency, and that HR7633 displays departures already on the second lobe indicates that this star is most likely the most asymmetric of the three. \\
Moreover, the closure phase departures of the three stars seem to be correlated with $\log{g}$ of the stars (Table~\ref{startable}), the latter is an indicator of the evolutionary phase: the largest deviations are for HR7633, which has $\log{g}=1.70$. This denotes that the size of the  granules become more significant with respect to the disk size as the surface gravity decreases, and therefore brightness fluctuations become more important. This idea is supported by previous work showing even larger departures from centrosymmetry for very evolved stars such as AGBs \citep{2016A&A...587A..12W,2010A&A...511A..51C,2006ApJ...652..650R} and red supergiant stars \citep{2016A&A...588A.130M,2010A&A...515A..12C}. However, we also note that we detected the largest deviations for the faintest star (HR7633), and this may indicate that we underestimated the errors.\\

%\subsection{Tentative of explanation of what is causing these departures?}

We now present a tentative explanation of these closure phase departures. A more complete analysis will be presented in a forthcoming paper.\\
Stellar surface asymmetries in the brightness distribution can be either due to convection-related and/or activity-related structures, to a companion, or to a clumpy dust envelope around the stars. In the following, we analyse the different possibilities.

A first hypothesis concerns convection-related surface structures affecting the interferometric observables. The expected convection turnover time in such a star is between hours to days or weeks, depending on the stellar fundamental parameters.  \cite{2010A&A...524A..93C,2014A&A...567A.115C} showed that stellar granulation manifests itself as surface asymmetries in the brightness distribution, and more precisely, in the closure phase signal. 

A second hypothesis is the stellar magnetic activity. \cite{2014A&A...567A.115C} and \cite{2015A&A...574A..69L} showed that star spots caused by the stellar magnetic field affect the closure phase signal in a similar way as the granulation. To determine its impact, we estimated the indicator S$_{\rm MW}$ based on the historic stellar activity data of chromospheric line emission. It measures the strength of the chromospheric emission core of the H and K lines of Ca II \citep{1982A&A...107...31M}. For the three stars, we converted S$_{\rm MW}$ into $\ensuremath{\log R^{\prime}_{\mathrm{HK}}}$ (Table~\ref{activity}) value using the calibration provided by \cite{2013A&A...549A.117M}  because $\ensuremath{\log R^{\prime}_{\mathrm{HK}}}$ is independent of the stellar type. In particular, we adopted their Eq.~13 and calibration for giant stars (their Eq.~23), which is valid for stars with 0.76 < B-V < 1.18. It should be noted that the B-V colour of HR7633 lies beyond the range of validity of the calibration, which implies that the activity indicator may not be reliable. The values for $\beta$ Oph and $\epsilon$ Cyg fall into or very close to the stellar inactivity region \citep[Figs.~3 and 4 in][]{2004AJ....128.1273W}. However, the star spots may still exist even if the activity is low, and their signature can only be distinguished by performing more interferometric observations  coupled with spectroscopic observations. \\
The source HR7633 appears to be an active star with $\ensuremath{\log R^{\prime}_{\mathrm{HK}}}=-4.132$. In addition to this, the stellar diameter of 178.11 R$_\odot$ of HR7633 (Table~\ref{phase}) as well as its low surface gravity (Table~\ref{startable}) seem to indicate that this star approaches the red giant tip towards the AGB phase. This evolutionary step is characterised by prominent stellar granulation accompanied by non-negligible dynamics \citep{2012A&A...547A.118L}. 

\begin{table}
%\begin{minipage}[t]{\columnwidth}
\caption{Photometric colours and stellar activity indicator of the observed RGB stars.}             % title of Table
\label{activity}      % is used to refer this table in the text
\centering                          % used for centering table
\renewcommand{\footnoterule}{} 
\begin{tabular}{c c c  c c  }        % centered columns (4 columns)
\hline\hline                 % inserts double horizontal lines
 Star & B-V & Parallax\tablefootmark{c} &  S$_{\rm MW}$\tablefootmark{d} & $\ensuremath{\log R^{\prime}_{\mathrm{HK}}}$ \\
         &       &          [mas]  &      &      \\
\hline
$\epsilon$ Cyg  & 1.04\tablefootmark{a} & 44.86$\pm$0.12 & 0.10  & -4.910 \\
HR7633 &  1.55\tablefootmark{a}  & 3.56$\pm$0.21 & 0.29 & -4.132 \\
$\beta$ Oph  &  1.18\tablefootmark{b} & 39.85$\pm$0.17 & 0.11 & -4.738 \\
\hline                        % inserts single horizontal line
 
\hline                                   %inserts single line
\end{tabular}
%\end{minipage}
\tablefoot{
\tablefoottext{a}{\cite{2002yCat.2237....0D}}
\tablefoottext{b}{\cite{1993A&AS..100..591O}}
\tablefoottext{c}{\cite{2007A&A...474..653V}}
\tablefoottext{d}{\cite{1991ApJS...76..383D}}
}
\end{table}

Concerning any possible companions, only $\epsilon$ Cyg is a known double system \citep{1994RMxAA..28...43P} where the primary star ($\epsilon$ Cyg) is $\approx25000$ brighter than its companion (dwarf M4 star) with $m_{V}$=2.45 \citep[$m_{H}$=0.20,][]{2002yCat.2237....0D} and $m_{V}$=11.98 ($m_{H}$=9.58\footnote{estimate based on PHOENIX models \citep[e.g. ][ and perso.ens-lyon.fr/france.allard/]{1997ARA&A..35..137A}}), respectively. In addition to this, the two stars lie 78.1$^{\arcsec}$ from each other. For these reasons, we estimate that the secondary star has a negligible effect on the observed data.

No dust production is detected around RGBs for $\beta$ Oph and HR7633 \citep[][ and McDonald private communication]{2011ApJS..193...23M}. The same is expected for $\epsilon$ Cyg. Dust is a sign of strong mass-loss; in evolved AGB stars, for example, the mass loss can be 1000 greater than in RGB stars.

Stellar convection-related and/or the magnetic-related surface activity are the most plausible explanation. To firmly establish the cause, it is necessary to monitor these stars with new interferometric observations, possibly coupled with spectropolarimetric data in the visible in order to measure their magnetic activity strength.

\section{Conclusions}

We presented observations of three RGB stars using the MIRC instrument at the CHARA interferometer. We showed that for all stars, the limb-darkening fit is very good 
with larger residuals for HR7633. We measured the apparent diameters of HD197989 ($\epsilon$ Cyg) = 4.61$\pm$0.02 mas, HD189276 (HR7633) = 2.95$\pm$0.01 mas, and HD161096 ($\beta$ Oph) = 4.43$\pm$0.01 mas.\\
Moreover, the closure phases denote departure points from the centrosymmetric case (closure phases not equal to 0 or $\pm180^{\circ}$) with values greater than 3$\sigma$. The departures seem to be qualitatively correlated with $\log{g}$ of the observed stars. HR7633, with the lowest $\log{g}$ of the sample, shows the highest deviations: the more the star evolves, the more significant
 the size of the  granules becomes with respect to the disk size.

We explored the possible causes of the break in symmetry in the brightness distribution and found that a possible explanation of this interferometric signal is the granulation and/or the stellar magnetic activity at its surface. However, it is not possible to confirm this at the moment, and a more complete analysis will be presented in a forthcoming paper.

\begin{acknowledgements}   

This work is based upon observations obtained with the
Georgia State University Center for High Angular
Resolution Astronomy Array at Mount Wilson Observatory.
The CHARA Array is supported by the National Science
Foundation under Grants No. AST-1211929 and AST-1411654.
Institutional support has been provided from the GSU
College of Arts and Sciences and the GSU Office of the
Vice President for Research and Economic Development. \\
AC, MM, PK, GP acknowledge financial support from "Programme National de Physique Stellaire" (PNPS) of CNRS/INSU, France. \\
RN acknowledges the Fizeau exchange visitors program in optical interferometry - WP14  OPTICON/FP7 (2013-2016, grant number 312430). \\
RC acknowledges the funding provided by The Danish National Research Foundation (Grant DNRF106). 
\end{acknowledgements}

%-------------------------------------------------------------------

   \bibliographystyle{aa}
\bibliography{biblio}

\begin{appendix}

\section{Additional figures for $\epsilon$ Cyg and HR7633}\label{appendix}

\begin{figure}
   \centering
   \begin{tabular}{c}                                                 
                                \includegraphics[width=0.95\hsize]{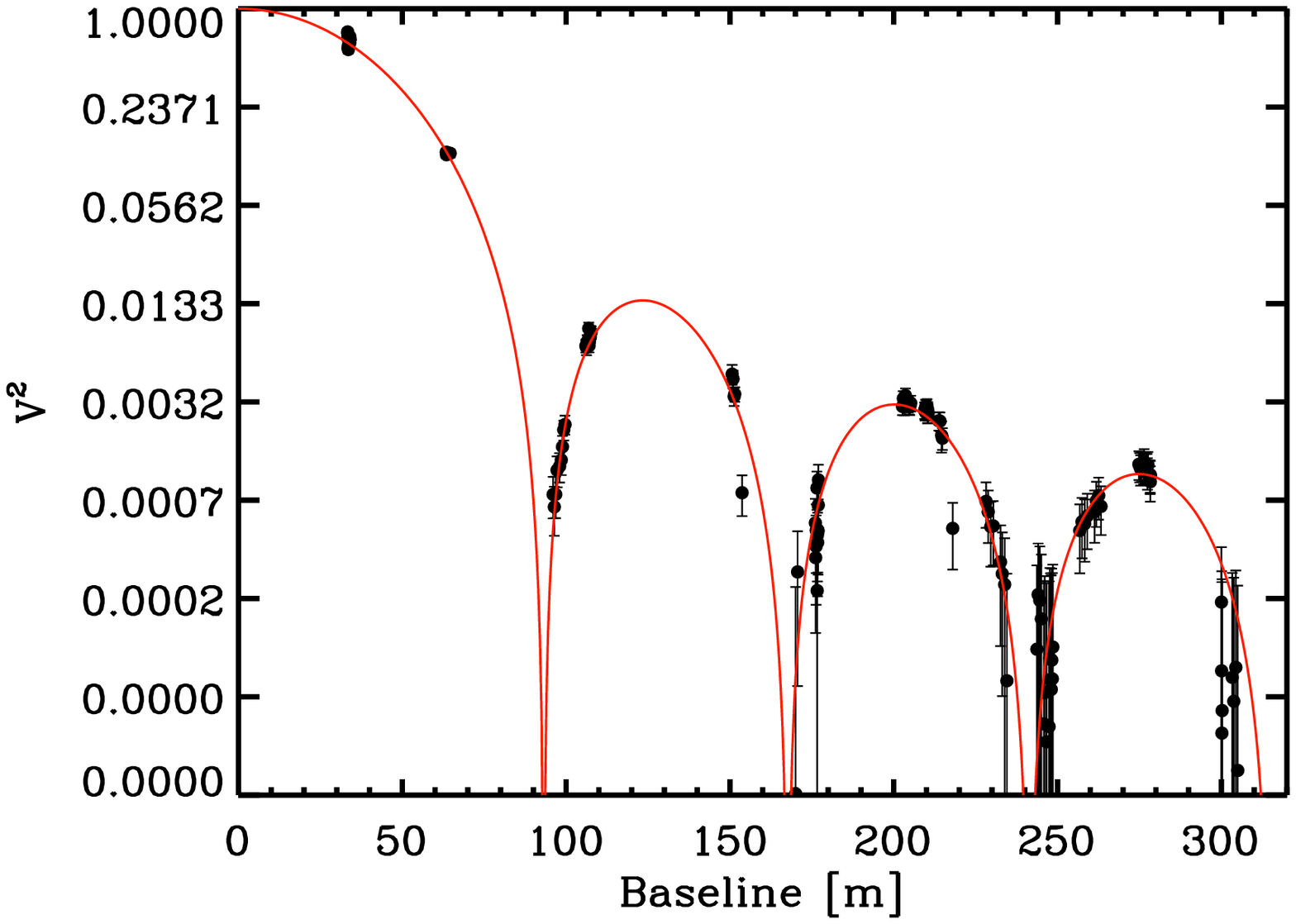}    \\
                                \includegraphics[width=0.935\hsize]{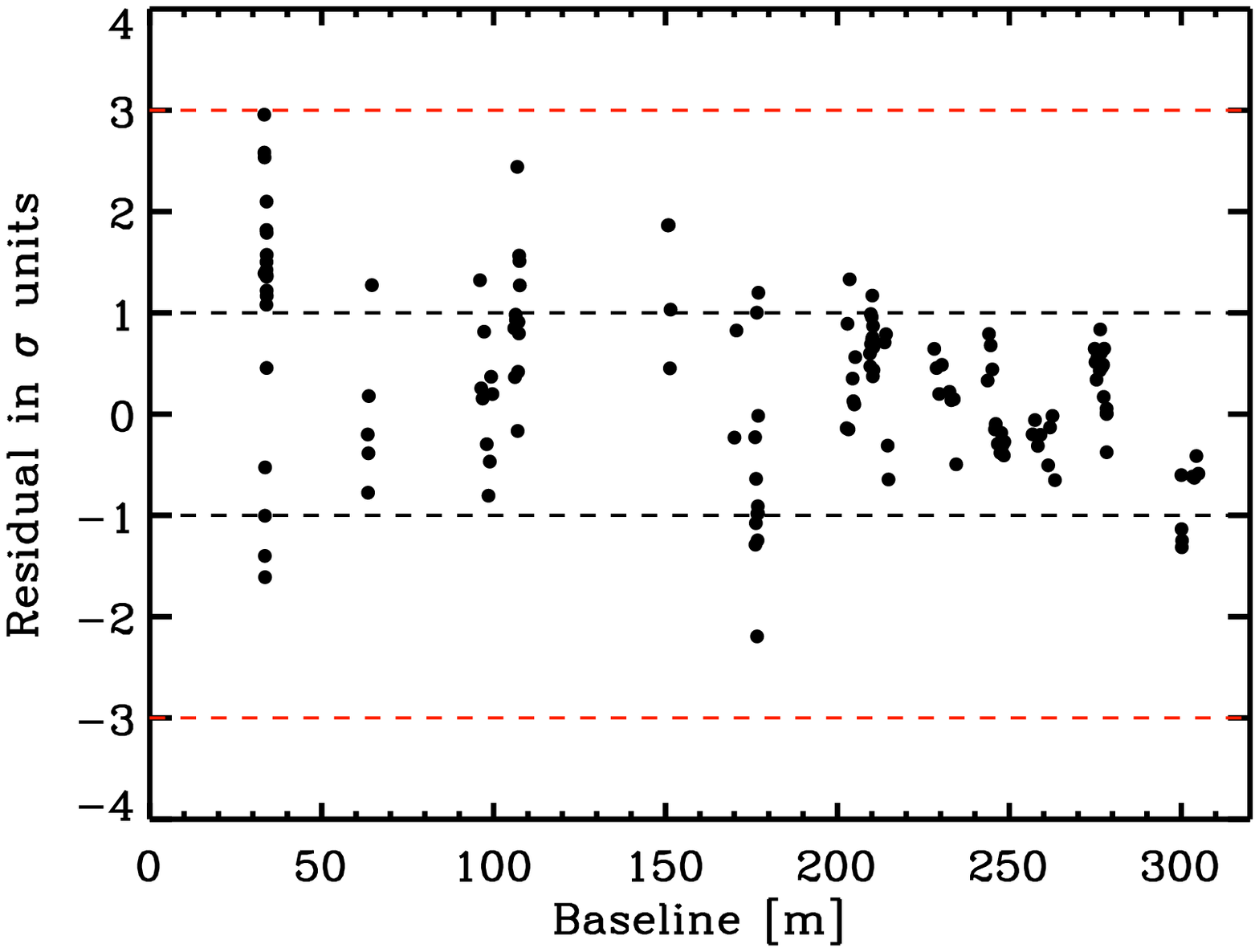}    \\
                             \includegraphics[width=0.95\hsize]{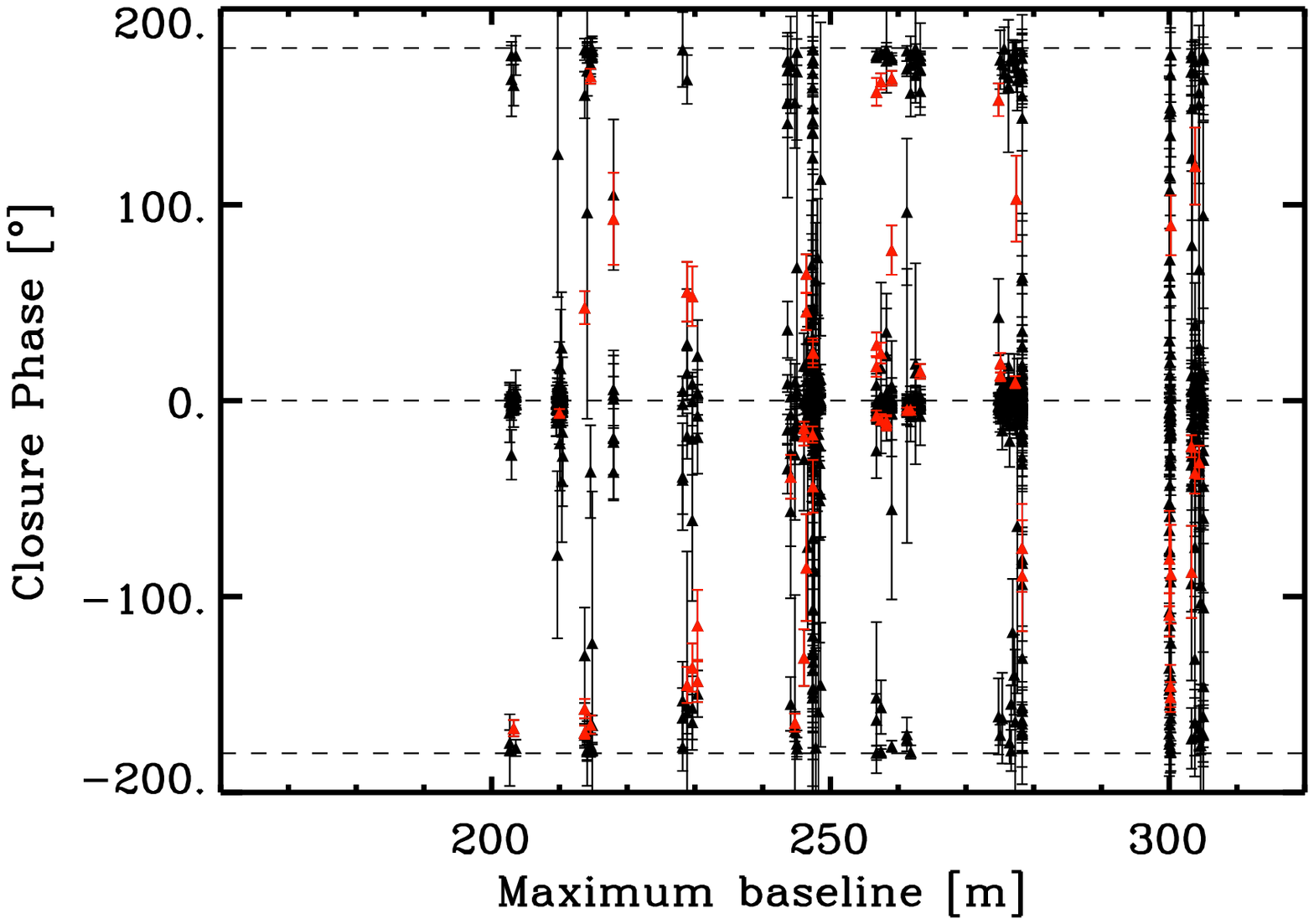}     
        \end{tabular}
      \caption{Same as in Fig.~\ref{betoph}, but for the star $\epsilon$ Cyg.}
        \label{epscyg}
   \end{figure}

\begin{figure}
   \centering
   \begin{tabular}{c}  
                                \includegraphics[width=0.95\hsize]{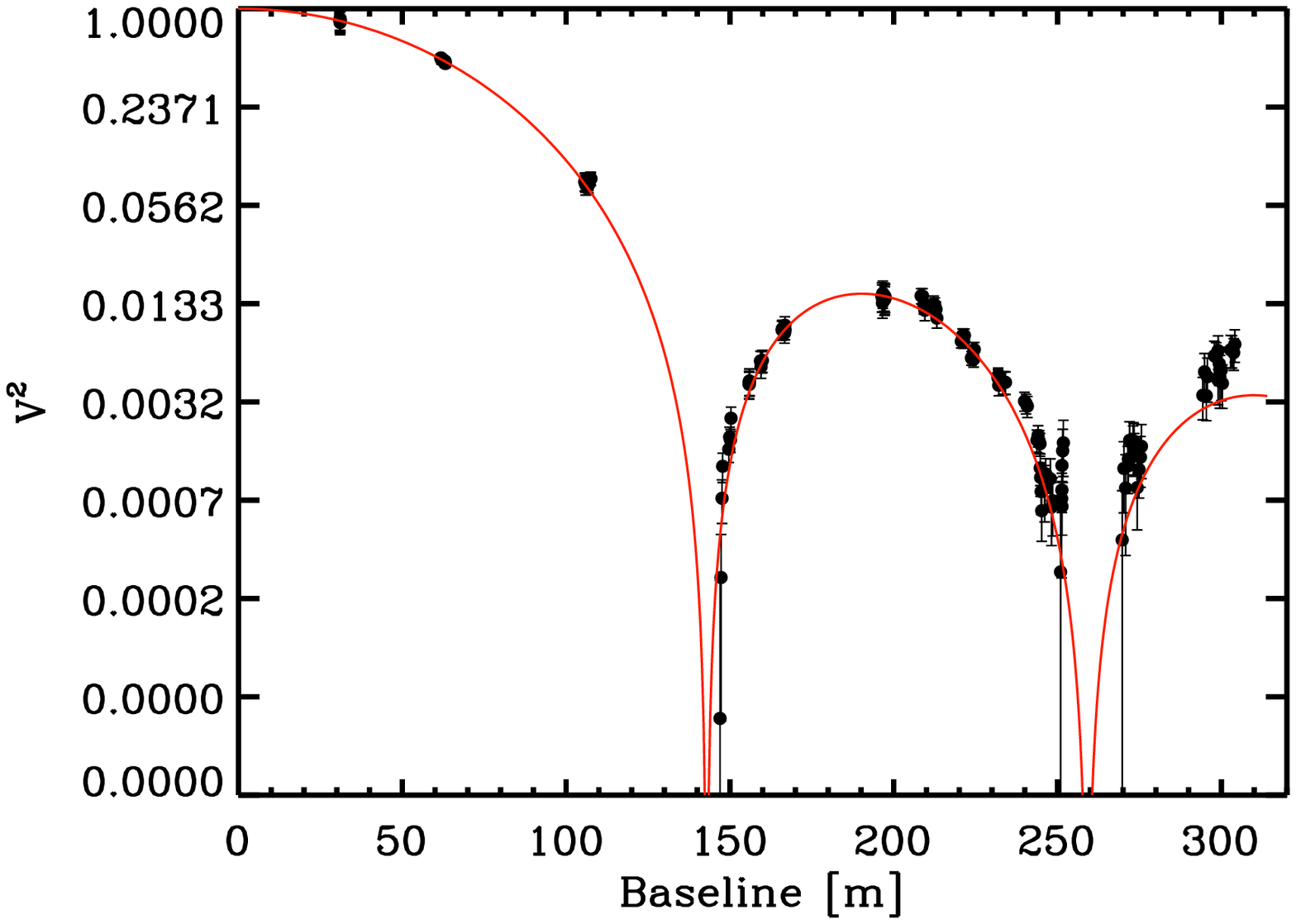}    \\
                                \includegraphics[width=0.935\hsize]{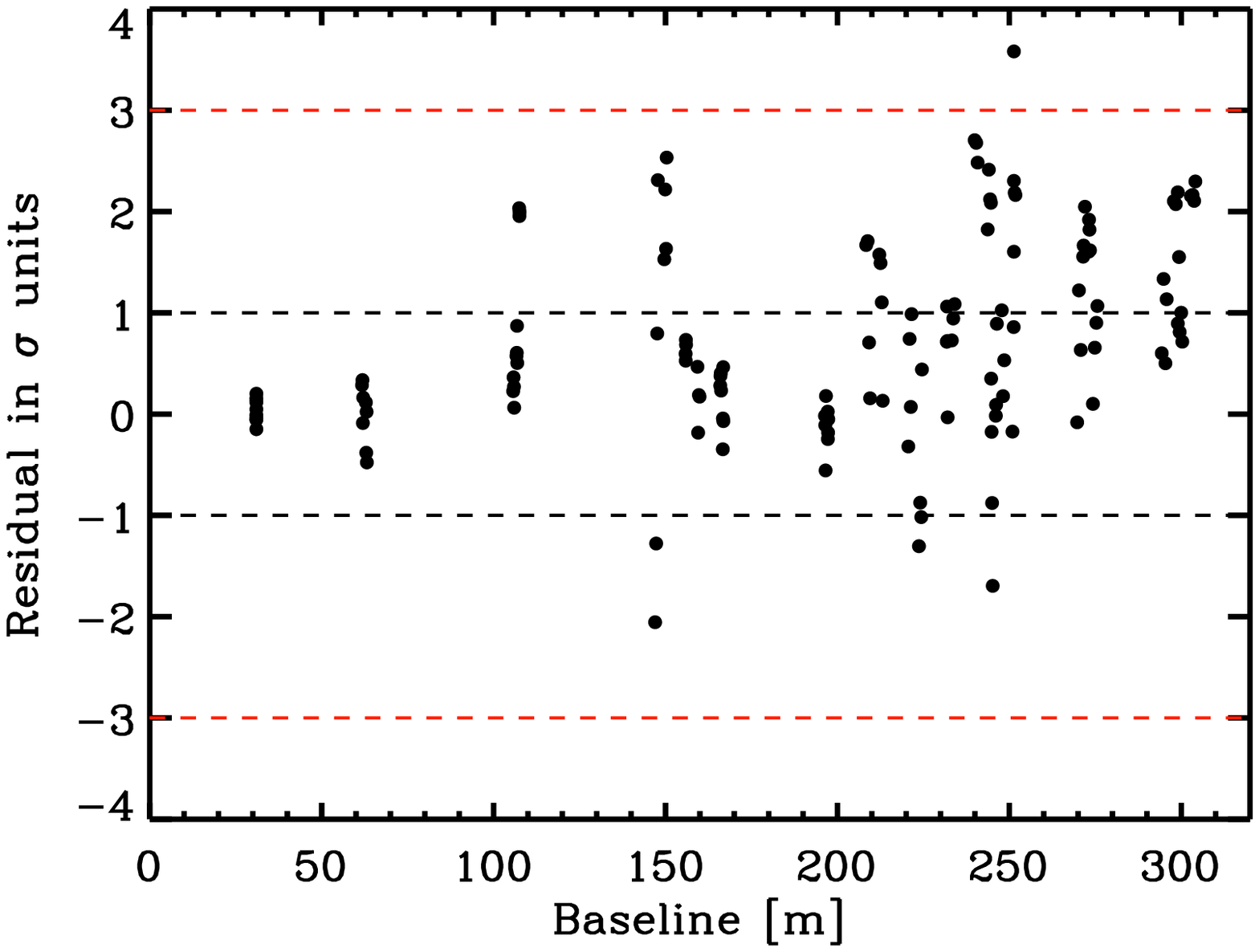}    \\
                             \includegraphics[width=0.95\hsize]{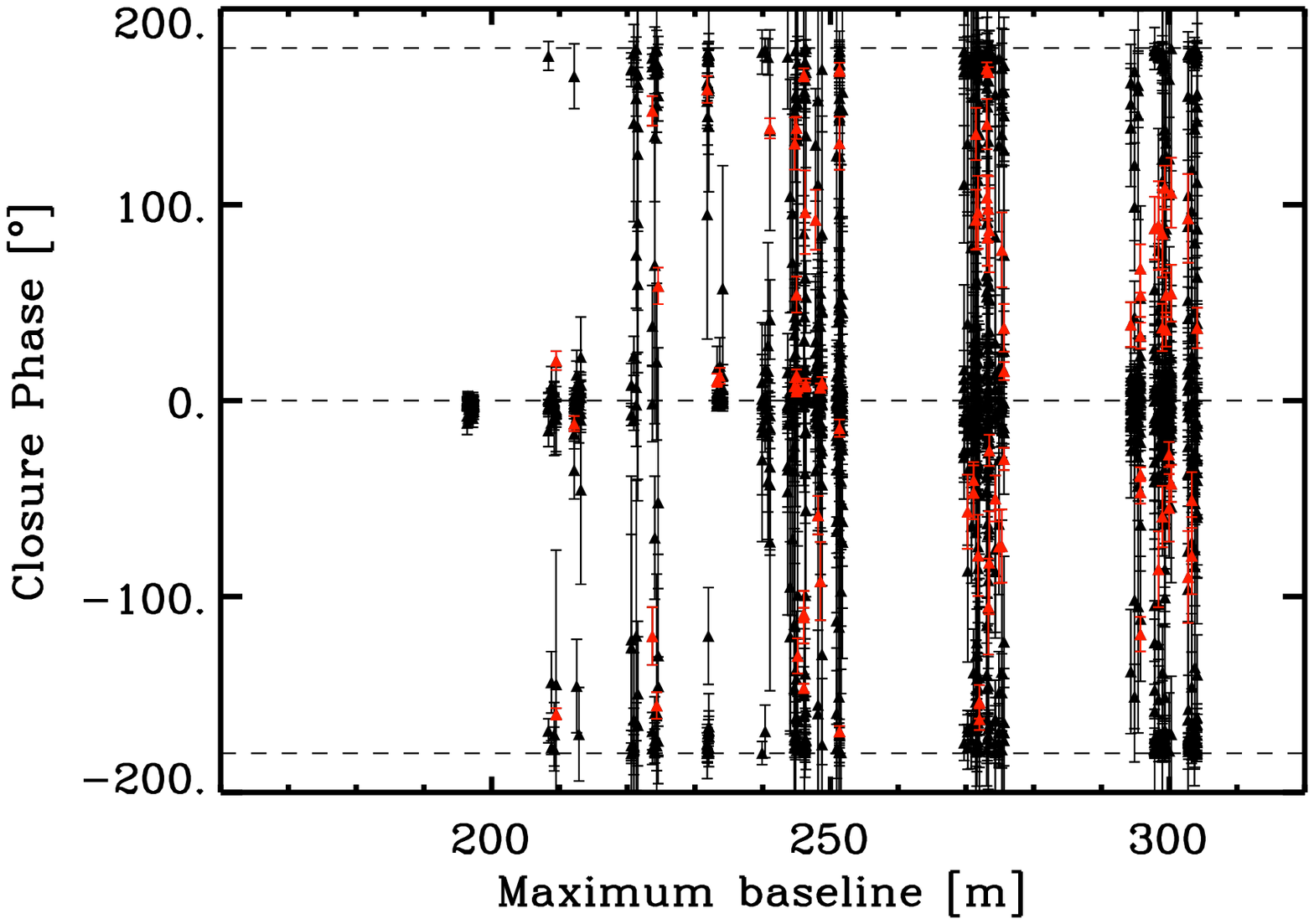}         
        \end{tabular}
      \caption{Same as in Fig.~\ref{betoph}, but for the star HR7633.}
        \label{HR7633}
   \end{figure}

\end{appendix}

\end{document}